\documentclass[aps,prb,superscriptaddress,twocolumn,bibliography,floatfix,10pt]{revtex4-2}
\usepackage[utf8]{inputenc}
\usepackage{amsfonts}
\usepackage{amssymb}
\usepackage{amsmath}
\usepackage{mathtools,units,xfrac,bbold,bm}
\usepackage{mathptmx}
\usepackage{graphicx}
\usepackage{cancel}
\usepackage[dvipsnames,svgnames]{xcolor}
\usepackage[normalem]{ulem}
\usepackage{soul} 
\usepackage{hyperref}
\hypersetup{colorlinks=true,citecolor=blue,linkcolor=blue,filecolor=blue,urlcolor=blue,pdfstartview=FitH,pdfpagemode=UseNone,pdfauthor={},pdfdisplaydoctitle=true,pdfduplex=DuplexFlipLongEdge}

\emergencystretch=\maxdimen
\newcommand{\figref}[2]{\hyperref[#1]{\ref*{#1}#2}}

\DeclareMathOperator{\Tr}{Tr}
\renewcommand{\rm}[1]{\mathrm{#1}}
\renewcommand{\bf}[1]{\mathbf{#1}}
\newcommand{\rb}[1]{\mathrm{\mathbf{#1}}}
\newcommand{\ave}[1]{\langle #1 \rangle}

\newcommand{\kv}{\mathbf k}
\newcommand{\rv}{\mathbf r}

\begin{document}
\normalem

\title{
	Finite temperatures and flat bands: the Hubbard model on three-dimensional Lieb lattices
}
\author{Lucas O. Lima}
\email{lucaslima@if.ufrj.br}
\affiliation{Instituto de F\'\i sica, Universidade Federal do Rio de Janeiro, Cx.P. 68.528, 21941-972 Rio de Janeiro RJ, Brazil}

\author{Juli\'an Fa\'undez}
\affiliation{Instituto de F\'\i sica, Universidade Federal do Rio de Janeiro, Cx.P. 68.528, 21941-972 Rio de Janeiro RJ, Brazil}
\affiliation{Departamento de Física y Astronomía, Universidad Andres Bello, Santiago 837-0136, Chile}

\author{Natanael C.~Costa}
\affiliation{Instituto de F\'\i sica, Universidade Federal do Rio de Janeiro, Cx.P. 68.528, 21941-972 Rio de Janeiro RJ, Brazil}

\author{Raimundo R.~dos Santos}
\affiliation{Instituto de F\'\i sica, Universidade Federal do Rio de Janeiro, Cx.P. 68.528, 21941-972 Rio de Janeiro RJ, Brazil}


%
%
\begin{abstract}
%
We investigate some thermodynamic and magnetic properties of the Hubbard model on two three-dimensional extensions of the Lieb lattice: the perovskite Lieb lattice (PLL) and the layered Lieb lattice (LLL). 
Using determinant quantum Monte Carlo (DQMC) simulations alongside Hartree-Fock and cluster mean-field theory (CMFT) approaches, we analyze how flat-band degeneracy, connectivity, and lattice anisotropy influence the emergence of magnetic order. 
Our results show that both geometries support finite-temperature magnetic transitions, namely ferromagnetic (FM) on the PLL, and antiferromagnetic (AFM) on the LLL.
Further, we have established that the critical temperature, $T_c$, as a function of the uniform on-site coupling, $U$, displays a maximum, which is smaller in the AFM case than in the FM one, despite the absence of flat bands in the LLL.  
We also provide numerical evidence to show that flat bands in the PLL rapidly generate magnetic moments, but a small interorbital coordination suppresses the increase of $T_c$ at large interaction strength $U/t$. 
By contrast, the LLL benefits from higher connectivity, favoring magnetic order even in the absence of flat bands. 
The possibilities of anisotropic interlayer hoppings and inhomogeneous on-site interactions were separateley explored.
We have found that magnetism in the PLL is hardly affected by hopping anisotropy, since the main driving mechanism is the preserved flat band; for the LLL, by contrast, spectral weight is removed from $d$-sites, which increases $T_c$ more significantly. 
At mean-field level, we have obtained that setting $U=0$ on $p$ sites and $U=U_d\neq0$ on $d$ sites leads to a quantum critical point at some $U_d$; this behavior was not confirmed by our DQMC simulations.
\end{abstract}

\maketitle

\section{Introduction}
\label{sec:introduction}
%
The interplay between lattice geometry and electronic band structure gives rise to fascinating quantum phenomena. 
In particular, the association of flat (or dispersionless) bands (FB's) \cite{leykam18,Deng2003,leykam24} with strong correlations and/or disorder has generated a wealth of unexpected physical properties.
Indeed, as a result of FB's favoring highly degenerate states one may find completely localized states at low cost of kinetic energy, thus leading to ferrimagnetism~\cite{Lieb89,Lieb89err,Costa16,Costa18}, room-temperature
ferromagnetism~\cite{mielke91,jiang2019lieb,bouzerar2023}, Mott physics~\cite{derzhko2015strongly,Po2018SUC,wu2007flat}, nontrivial topological states~\cite{Weeks10,jiang2019topology,Tang2011FQHS,Sun2011topology} and enhanced electronic correlations~\cite{andrade2023topological}.

\begin{figure}[t] 
\includegraphics[width=\linewidth]{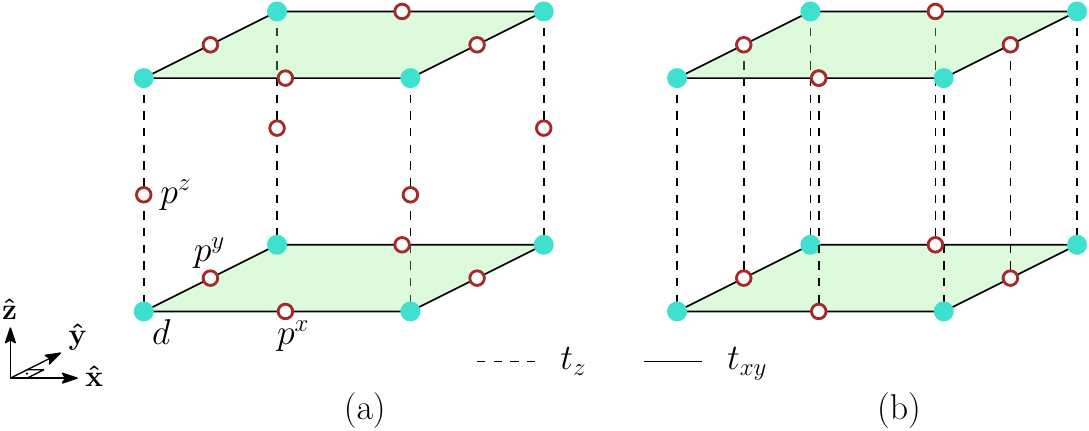}
\caption{
3D extensions of the Lieb lattice: (a) as a perovskite lattice (PLL) and (b) as stacked layers (LLL). The unit cell in the LLL configuration is identical to that of the 2D case, consisting of $ d $, $ p^x $, and $ p^y $ sites in the $ xy $ plane. In the PLL model, an additional $ p $ site is introduced along the $ z $-axis, with $ d $, $ p^x $, $ p^y $, and $ p^z $ sites forming the unit cell. Solid and dashed lines denote hopping amplitudes in the $ xy $-plane, $ t_{xy} $, and along the $ z $-direction, $ t_z $, respectively.
}
\label{fig:lattice}
\end{figure}

Amongst various two-dimensional (2D) geometries capable of hosting FB's \cite{dias2015}, the Lieb lattice (LL) has emerged as a fertile testing ground to study interacting electrons. 
Also referred to as the CuO$_2$, or decorated square lattice, the LL features a unit cell formed by three sites, usually referred to as $d$, $p^x$, and $p^y$ sites [see the $xy$ planes in Fig.\,\ref{fig:lattice}].
A tight-binding treatment of electrons hopping between nearest neighbor sites of this geometry yields highly localized states at $p$ (or O) sites~\cite{GuzmanSilva14, Vicencio15, Slot17}, which become occupied at half hilling. 
As a result, when an on-site repulsion, $U$, is switched on on every site, the ground state becomes ferrimagnetic~\cite{Lieb89,Lieb89err,Costa16}.
Due to the continuous symmetry of the order parameter, ferrimagnetism is unstable at any finite temperatures.

If one wants to investigate the effects brought about by FB's at finite temperatures, one possible scenario is to stack two-dimensional LL's along the direction perpendicular to the CuO$_2$ plane. 
From a theoretical perspective, here it suffices to consider two possibilities of stacking the LL's, as shown in Fig.\,\ref{fig:lattice}.
We may setup a perovskite-like lattice [Fig.\,\figref{fig:lattice}{(a)}], in which $p$ sites are introduced in-between $d$ sites of successive LL planes, so that the resulting lattice has cubic symmetry with each face forming a 2D LL; we refer to this as the perovskite Lieb lattice (PLL).
Alternatively, we may simply pile up LL planes, in such way that vertical hopping between successive planes are allowed between $d$ sites and between $p$ sites [Fig.\,\figref{fig:lattice}{(b)}]; we refer to this as the layered Lieb lattice (LLL).

%
%

These two forms of stacking LL layers lead to different numbers of sites in their unit cells (see Fig.\,\ref{fig:lattice}), so that they are expected to exhibit distinct physical properties, particularly in relation to the retention of a FB at the Fermi level, $\epsilon_F$.
While the three-dimensional (3D) PLL preserves the doubly degenerate FB at $\epsilon_F$ for half filling \cite{Weeks10}, the FB is suppressed for the LLL \cite{Noda14}. 
The presence or absence of FB's plays a critical role in shaping the many-body physics of these systems, including the emergence of magnetic phases and the evolution of the critical transition temperature $T_c$, which depends on the density of states $\rho(\epsilon_F)$ and the strength of the fermionic on-site interaction $U$. 
Exploring correlation effects in these two distinct extensions of the LL to three dimensions may thus offer new insights into the physics of FB systems. 
For completeness, we should mention that the synthesis of materials exhibiting 3D FB's has recently been achieved\,\cite{wakefield2023three,Huang_2024}.

While a great wealth of data, both numerical \cite{scalettar89, kent05, kozik13, staudt00, paiva11, hirschmeier15} and experimental \cite{imrivska14, hart15}, has been accumulated about the half-filled Hubbard model on the simple cubic lattice, much less is known about 3D extensions of LL's.
A LLL stacking of Hubbard LL's with a finite number $L$ of layers was considered in Ref.\,\cite{Noda14}, a scenario for which Lieb's theorem\,\cite{Lieb89,Lieb89err} predicts the onset of ferromagnetism (FM) or antiferromagnetism (AFM) in the ground state, respectively for even or odd $L$. 
Further DMFT work at finite temperatures \cite{Noda15} concentrated on anisotropic hoppings to isolate the effects of van Hove singularities and of FB's on the LLL's; the qualitative behavior of $T_c(U)$ for antiferromagnetism is similar to that for the simple cubic lattice.

In view of the differences brought about by the presence or absence of FB's, a direct comparison of finite-temperature properties of the half-filled Hubbard model on these two 3D LL extensions offers an excellent framework to assess how the spectral features influence magnetic ordering and critical behavior.
To this end, we use determinant quantum Monte Carlo (DQMC) simulations, which provide unbiased results for strongly correlated fermionic systems. 
Furthermore, we stress that the interest in the study of such lattice geometries is not restricted to materials, but also extends to highly tunable platforms such as ultracold atoms in optical lattices\,\cite{kohl2005, jordens2008, schneider2008, Noda09, Esslinger10, Noda14, Taie15, Duarte2015, hart15, schafer2020}, where flat-band physics and Hubbard-like interactions can be realized and explored experimentally.

The layout of the paper is as follows. 
In Sec.\,\ref{sec:model}, we present the model and the main features of the DQMC method. 
In Sec.\,\ref{sec:results}, we examine the main properties of the noninteracting case, and analyze thermodynamic and magnetic properties of the interacting case. 
We supplement DQMC data with Hartree-Fock (HF) calculations for weak coupling, and with cluster mean field theory (CMFT) at strong coupling.
Section \ref{sec:conclusions} summarizes our findings.

\section{Model and methodology}
\label{sec:model}

The Hubbard Hamiltonian for 3D extensions of the Lieb lattice may be written as
\begin{equation}\label{eq:Hamiltonian}
\widehat{\mathcal{H}} = \widehat{H}_K + \widehat{H}_U + \widehat{H}_\mu,
\end{equation}
where $\widehat{H}_K = \widehat{H}_{xy} + \widehat{H}^\gamma_{z}$ denotes the kinetic energy for motion along the $ xy $-plane and $ z $-direction, respectively, and  $\gamma$ stands for PLL or LLL; $ \widehat{H}_U $ describes the on-site interaction and $ \widehat{H}_\mu $ controls the band filling. 
These terms are given by
\begin{subequations}
\begin{align}
\label{eq:kinetic_energy}
\widehat{H}_{xy} = & -t_{xy}\, \sum_{\mathbf{r},\sigma}\left( d^{\dagger}_{\mathbf{r},\sigma} p^x_{\mathbf{r},\sigma} + d^{\dagger}_{\mathbf{r},\sigma} p^y_{\mathbf{r},\sigma} + \mathrm{H.c} \right) \nonumber \\
& -t_{xy}\, \sum_{\mathbf{r},\sigma}\left( p^{x\dagger}_{\mathbf{r},\sigma} d_{\mathbf{r} + \hat{\rb{x}},\sigma} + p^{y\dagger}_{\mathbf{r},\sigma} d_{\mathbf{r} + \hat{\rb{y}},\sigma} + \mathrm{H.c} \right), \\
\label{eq:Hz_PLL}
\widehat{H}^\text{PLL}_{z} = & -t_z \, \sum_{\mathbf{r},\sigma}\left( d^{\dagger}_{\mathbf{r},\sigma} p^{z}_{\mathbf{r},\sigma} + p^{z\dagger}_{\mathbf{r},\sigma} d_{\mathbf{r} + \hat{\rb{z}},\sigma} + \mathrm{H.c} \right), \\
\label{eq:Hz_LLL}
\widehat{H}^\text{LLL}_{z} = & -t_z \, \sum_{\mathbf{r},\sigma}\left( d^{\dagger}_{\mathbf{r},\sigma} d_{\mathbf{r} + \hat{\rb{z}},\sigma} + p^{x\dagger}_{\mathbf{r},\sigma} p^{x}_{\mathbf{r} + \hat{\rb{z}},\sigma} + p^{y\dagger}_{\mathbf{r},\sigma} p^{y}_{\mathbf{r} + \hat{\rb{z}},\sigma} + \mathrm{H.c} \right), \\
\label{eq:Potential_energy}
\widehat{H}_{U} = & ~ \sum_{\mathbf{r},\alpha} U_\alpha \left( \hat{n}^{\alpha}_{\mathbf{r},\uparrow} - 1/2 \right)\left( \hat{n}^{\alpha}_{\mathbf{r},\downarrow} - 1/2 \right), \\
\label{eq:Chemical_energy}
\widehat{H}_\mu = & -\mu \sum_{\mathbf{r},\sigma,\alpha} \hat{n}^{\alpha}_{\mathbf{r},\sigma}  \; ,
\end{align}
\end{subequations}
with $d^{\dagger}_{\mathbf{r},\sigma}$, $p^{x\dagger}_{\mathbf{r},\sigma}$, $p^{y\dagger}_{\mathbf{r},\sigma}$ and $p^{z\dagger}_{\mathbf{r},\sigma}$ being standard fermion creation operators acting on orbital $ \alpha =d$, $p^x$, $p^y$ or $p^z$ at position $\rb{r}$ with spin $\sigma$, while $ \hat{n}^{\alpha}_{\mathbf{r},\sigma} $ are the corresponding number operators.   
The two terms on the right-hand side of Eq.\,\eqref{eq:kinetic_energy} denote the intra- and intercell hopping between $ d $ and $ p $ orbitals in the $ xy $ plane, respectively. Equations \eqref{eq:Hz_PLL} and \eqref{eq:Hz_LLL} describe the hopping along the $z$-axis and between orbitals in the $xy$ plane for the PLL and LLL, respectively. $U_\alpha$ denotes the strength of the on-site Coulomb repulsion in a given orbital $\alpha$; unless otherwise mentioned, we assume $U_\alpha = U$ for all orbitals. Finally, $\mu$ represents the chemical potential, which controls the filling of the electronic states.
With the symmetrized definition of Eq.\,\eqref{eq:Potential_energy}, we set  $\mu=0$ to yield a half-filled band; we also set $ t_{xy} = t =1 $ as the energy scale.

The physical properties of the Hamiltonian $\mathcal{H}$, Eq.\,\eqref{eq:Hamiltonian}, are extracted through  DQMC simulations \cite{blankenbecler81,hirsch85,white89,dosSantos03b}. 
This method provides unbiased numerical solutions, mapping a $d$-dimensional interacting system onto an equivalent $(d+1)$-dimensional system with an additional imaginary-time dimension, $\beta \in [0,\tau]$, where $\beta$ is the inverse temperature, $T$; we also set the Boltzmann constant, $k_B$, as unity.  
In this approach, the one-body $\hat{\mathcal{K}}$ and two-body $\hat{\mathcal{P}}$ operators in the partition function, $\mathcal{Z}$, are separated using the Trotter-Suzuki (TS) decomposition. 
This is achieved by defining $\beta = l_{\tau} \Delta \tau$, with $l_\tau$ being the number of imaginary-time slices and $\Delta \tau$ as the time step. Thus,
\begin{align}
\label{eq:TS-decoupling}
\mathcal{Z} &= \Tr [e^{-\beta\widehat{\mathcal{H}}}] = \Tr\,[(e^{-\Delta\tau(\hat{\mathcal{K}} + \hat{\mathcal{P}})})^{l_\tau}] \nonumber \\ &\thickapprox \Tr\, [e^{-\Delta\tau\hat{\mathcal{K}}}e^{-\Delta\tau\hat{\mathcal{P}}}e^{-\Delta\tau\hat{\mathcal{K}}}e^{-\Delta\tau\hat{\mathcal{P}}} \cdots].
\end{align}
Equation~\eqref{eq:TS-decoupling} leads to an error proportional to $(\Delta \tau)^{2}$, which can be systematically reduced as $\Delta\tau\to 0$. Here, we choose $\Delta\tau\leq0.1$, which is small enough so that the systematic errors from the TS decomposition are comparable to the statistical ones (i.e., from the Monte Carlo sampling).
The next step is to perform a discrete Hubbard-Stratonovich (HS) transformation~\cite{Hirsch83} on the two-body terms, $\exp{(-\Delta\tau\hat{\mathcal{P}})}$, which converts them also to quadratic form in fermion operators, at the cost of introducing discrete HS auxiliary fields, $s(\rb{r},\tau)$. 
In this way the resulting trace of fermions propagating in an auxiliary bosonic field can be performed. 
Thus, one can evaluate Green's functions and other physical observables including spin-, charge- and pair correlation functions by sampling the HS fields with the product of fermionic determinants acting as Boltzmann weights.

Although the DQMC method is unbiased, in the absence of some specific symmetries \cite{li16}, such as particle-hole symmetry, it may suffer from the infamous minus-sign problem~\cite{Loh90,troyer2005computational,mondaini2022_2} at low-temperatures; for a detailed discussion, see, e.g., Refs.\,\cite{dosSantos03b,assaad02,gubernatis16,Becca17}.
We emphasize that here we only consider the half-filled band case, which leads to sign-free simulations.

Signatures of finite temperature phase transitions and long-range order are sought in the behavior of several quantities, such as the internal energy,
\begin{align}
\label{eq:energy}
e(T) &= \frac{1}{N}  \langle\,\widehat{\mathcal{H}}\,\rangle,
\end{align}
from which we obtain the specific heat per orbital,
\begin{align}
\label{eq:specific_heat}
c(T) &= \frac{1}{N} \frac{d\langle\, \widehat{\mathcal{H}} \,\rangle}{dT}.
\end{align}

For the magnetic response, we probe the local moment,
\begin{equation}
\langle m_{\alpha}^{2} \rangle = \langle (\hat{n}_{\mathbf{r}, \uparrow}^{\alpha} - \hat{n}_{\mathbf{r}, \downarrow}^{\alpha} )^{2} \rangle,
\label{eq:local_moment}
\end{equation}
as well as real space spin-spin correlation functions,
\begin{equation}
\label{eq:Spincorrelation}
c^{\alpha\alpha^\prime} (\bf{r})= \langle (\hat{n}_{\bf{r}_0, \uparrow}^{\alpha} - \hat{n}_{\bf{r}_0, \downarrow}^{\alpha} )(\hat{n}_{\bf{r}_{0} + \bf{r}, \uparrow}^{\alpha^\prime} - \hat{n}_{\bf{r}_{0} + \bf{r}, \downarrow}^{\alpha^\prime} )\rangle \, ,
\end{equation}
with $\bf{r}_{0}$ being the position of a given unit cell, while $\alpha$ and $\alpha^\prime$ denote the orbitals $d$, $p^x$, $p^y$ or $p^z$.
We also resort to the magnetic structure factor,
\begin{equation}
\label{eq:StructFactor}
S(\bf{q}) = \frac{1}{N} \sum_{\alpha\alpha^\prime}\sum_{\bf{r}}\,c^{\alpha\alpha^\prime}(\bf{r})\,e^{\rm{i}\,\bf{q}\cdot\bf{r}} \, ,
\end{equation}
where $N$ is the total number of lattice orbitals. 

With these structure factors, we set up correlation ratios \cite{Kaul15,darmawan18},
\begin{equation}
   R_{\eta}(\tilde{L}) = 1 - \frac{S_\eta \left(\bf{q} - \delta\bf{q}\right)}{S_\eta \left(\bf{q}\right)},
  \label{eq: Rc}
\end{equation}
where $\eta$ denotes the pertinent correlations probed, namely FM [$\bf{q}=(0,0,0)$] or AFM [$\bf{q}=(0,0,\pi)$], and $|\delta \bf{q}| = 2\pi/\tilde{L}$  represents the discrete momentum intervals for a lattice with linear dimension $\tilde{L}=(L_x,L_y,L_z)$. 
Being a dimensionless size-scaling invariant quantity, when plotted for different lattice sizes, the intersection of the $R_\eta$ curves provides estimates for the critical point; these, in turn, may be subsequently extrapolated towards the thermodynamic limit with the aid of finite-size scaling \cite{Fisher71,Barber83,dosSantos81a}.

In addition to DQMC simulations, we resort to HF calculations in the weak coupling regime ($U/t \ll 1.0$), and to CMFT for the strong coupling limit (\(U/t \gg 1.0\)); details of these approaches are provided in Appendices \ref{Ap} and \ref{CMFT}, respectively.

\section{Results}
\label{sec:results}
%
\begin{figure}[t] 
\includegraphics[width=\linewidth]{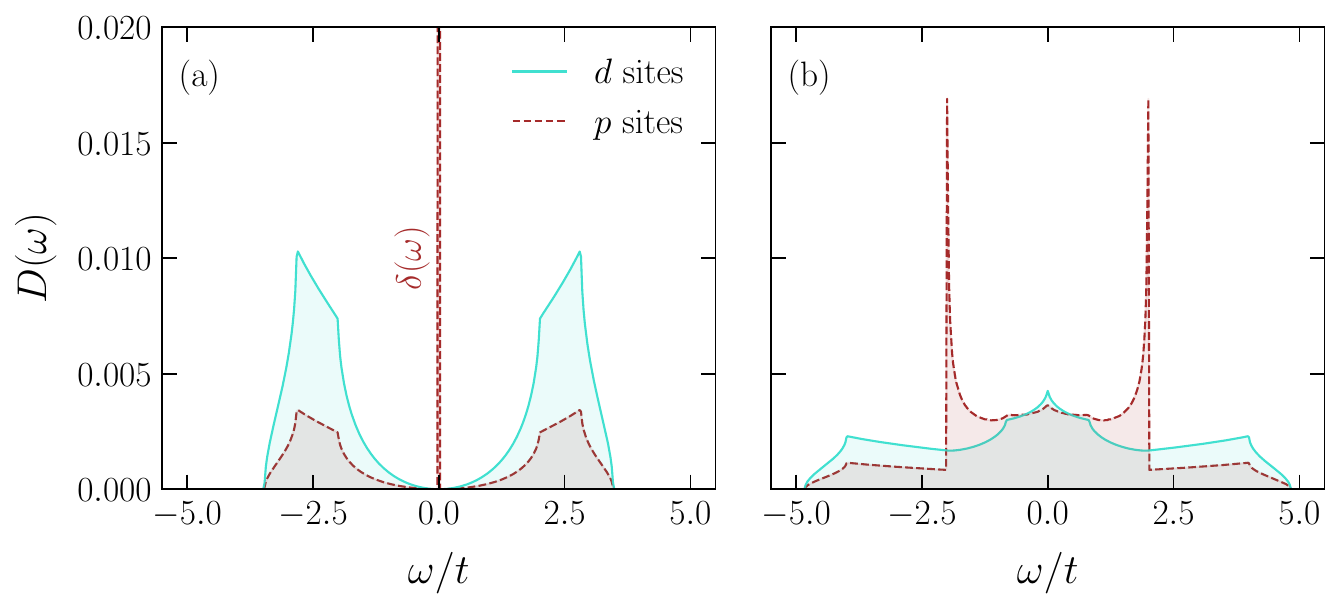}
\caption{
\label{fig:DOS} 
Site-resolved non-interacting DOS for tight-binding fermions on (a) PLL and (b) LLL. The energy $\hbar\omega$ is measured relative to the $\epsilon_F$, assuming half filling.
}
\end{figure}
%
\subsection{Non-interacting limit}
When $U/t=0$, $\widehat{\mathcal{H}}$ can be straightforwardly diagonalized in $\bf{k}$-space for both lattices. 
For the PLL, part of the energy spectrum consists of two dispersive bands, 
\begin{equation}
	\label{eq:bands_disp_PLL}
	\epsilon_{\pm}(\bf{k}) = \pm t \sqrt{2[\cos k_x + \cos k_y + \cos k_z + 3]},
\end{equation}
associated with the $d$ sites, lying symmetrically in energy with respect to $ \omega\equiv \epsilon(\kv)-\epsilon_F=0$; the corresponding DOS's are shown in Fig.\,\figref{fig:DOS}{(a)}. 
Given that the PLL shares the same symmetry of the 2D LL along the three Cartesian directions, the flat band associated with the $p$ sites is preserved, manifested by a two-fold degenerate $\delta$-function at $\omega=0$ in the DOS plot in Fig.\,\figref{fig:DOS}{(a)}.
That is, the presence of $p$ sites between planes contributes to localize electrons isotropically.

For the LLL stacking, on the other hand, no FB is formed; instead, one finds three dispersive bands,
\begin{subequations}
\begin{align}
\label{eq:bands12_disp_LLL}
\epsilon_{\pm}(\bf{k}) &= \pm\ t \sqrt{2[\cos k_x + \cos k_y + 2]} - 2t\cos k_z, \\
\label{eq:bands3_disp_LLL}
\epsilon(\bf{k}) &= - 2t\cos k_z,
\end{align}
\end{subequations}
whose corresponding DOS's are shown in Fig.\,\figref{fig:DOS}{(b)}. 
The absence of a FB in this case creates channels for electronic delocalization. 
Although flat bands are absent, a van Hove singularity appears at half-filling, as shown by the peaks in the DOS at $\omega=0$, in Fig.\,\figref{fig:DOS}{(b)}.
%
\begin{figure}[t] 
\includegraphics[width=\linewidth]{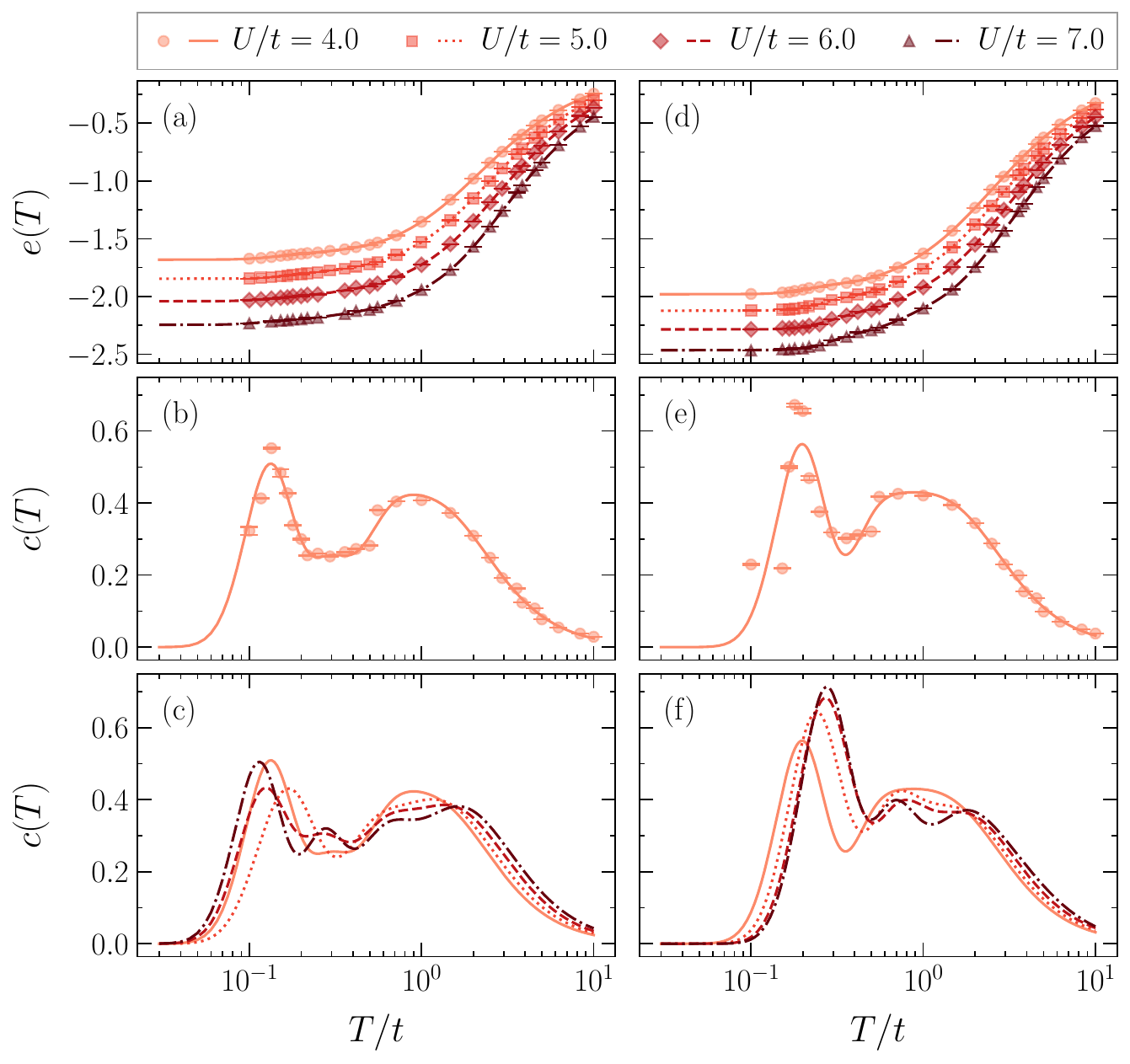}
\caption{
\label{fig:SHeat} 
 Internal energy and specific heat as a function of the temperature (linear-log scale) for various values of $U/t$ and fixed lattice size $\tilde{L}=(4,4,4)$, shown for PLL (left panels) and LLL (right panels), respectively. Panels (a) and (d): Symbols represent internal energy data from DQMC simulations, while lines show exponential fits using the function $f_{\text{fit}}$ (see text). Panels (b) and (e): Specific heat as a function of temperature for fixed $U/t=4.0$, with symbols indicating numerical differentiation of the DQMC internal energy data in panels (a) and (d), and lines showing differentiation of the fitted function $f_{\text{fit}}$. Panels (c) and (f): Specific heat as a function of temperature, derived from the full differentiation procedure of $f_{\text{fit}}$ for various values of $U/t$.
}
\end{figure}
%

\subsection{Specific heat data}
\label{subsec:Therm. Prop}
%
We proceed by analyzing the interacting case, $U/t > 0$. 
From this point onwards, the observables defined in Sec.\,\ref{sec:model} are investigated for lattice sizes $4 \times 4 \times 4$ ($N=n_s\times4^3$ sites, with $n_s$ being the number of orbitals within the unit cell), unless stated otherwise. 
We start by discussing the internal energy per particle, Eq.\,\eqref{eq:energy}, and the specific heat, Eq.\,\eqref{eq:specific_heat}.
Figure \ref{fig:SHeat} shows the behavior of $e(T)$ and $c(T)$ for different values of $U/t$, with panels on the left (right) column corresponding to data for the PLL (LLL). 
Figures \figref{fig:SHeat}{(a)} and \figref{fig:SHeat}{(d)} compare $e(T)$ for both lattices; 
solid lines correspond to the function $f_{\rm{fit}} = a_0 + \sum_{n=1}^{\le 9} a_n \exp(-n\beta\Delta)$, where the parameters $a_n$ and $\Delta$ are determined through least-squares fits to $e(T)$.
Figures \figref{fig:SHeat}{(b)} and \figref{fig:SHeat}{(e)} compare the specific heat for $U/t=4.0$, as determined from finite differences of the QMC points with those obtained from differentiating $f_\text{fit}$. 
We see that the specific heat obtained from the fitting adequately describes the structure of the peaks, at a much lower computational effort. 
Accordingly, Figures \figref{fig:SHeat}{(c)} and \figref{fig:SHeat}{(f)} show the specific heat data thus obtained for a wide range of values of $U/t$, which we now discuss.
As the temperature is decreased, the specific heat first displays a broad peak, which is typically associated with the formation of local moments \cite{Paiva2001}. 
Further decrease in the temperature leads to another, narrower peak, which, in three dimensions, signals ordering of the moments formed at higher temperatures.

While one is unable to distinguish the nature of the magnetic ordering from specific heat data alone, we may highlight some differences between the behaviors on the two lattices.  
First, intermediate and less intense peaks appear for $U/t \gtrsim 7.0$ on both lattices, which may indicate that local moments form on $d$ and $p$ sites at different temperatures in the strong coupling regime.
Second, for the LLL we note that the low-$T$ peak shifts to higher temperatures as $U/t$ increases; by contrast, for the PLL the low-$T$ peak moves towards lower temperatures.
Given the different scales of temperatures involved, one may therefore expect critical temperatures for magnetic ordering to be quite distinct for both lattices. 

%
\begin{figure}[b] 
\includegraphics[width=\linewidth]{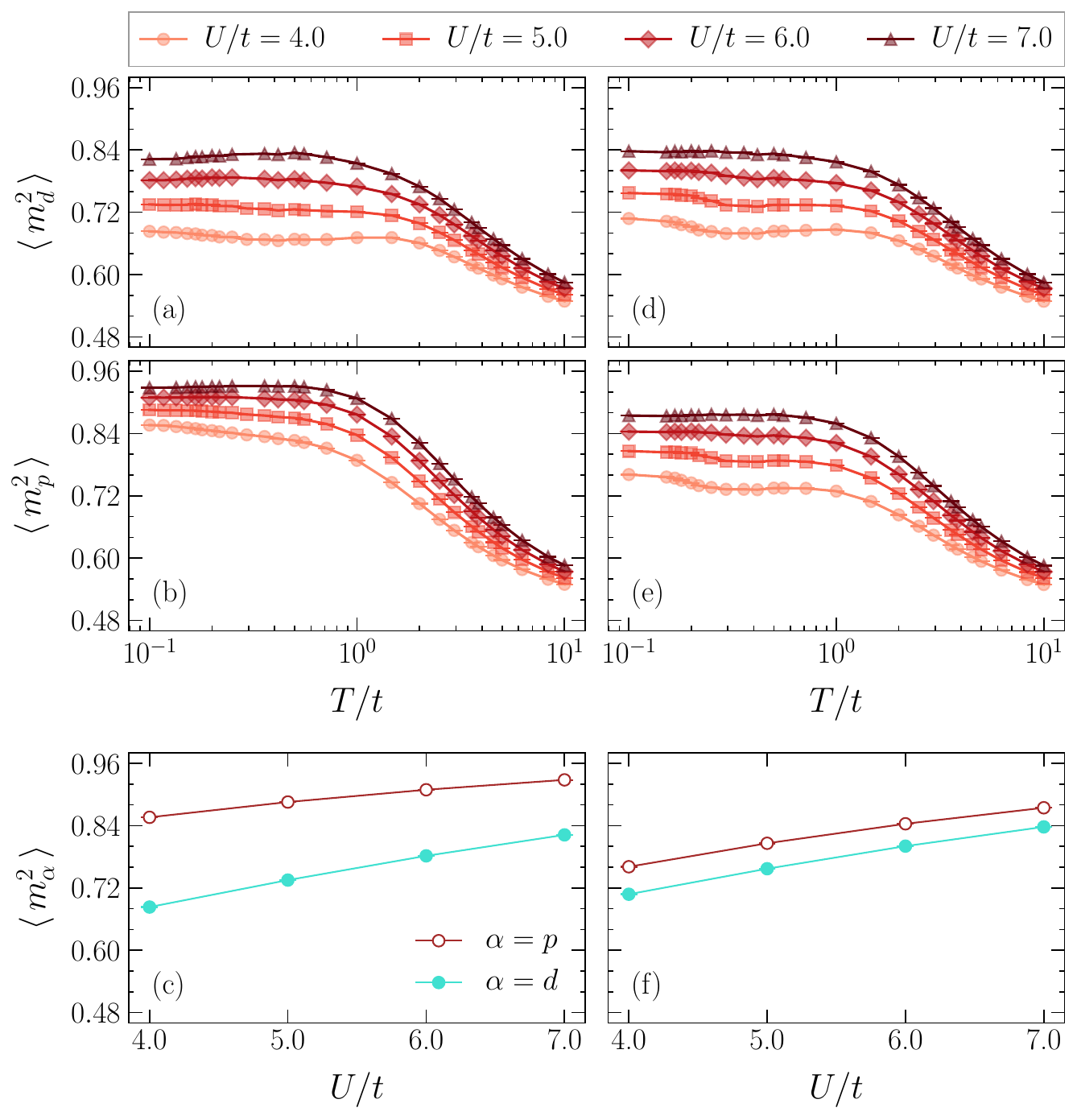}
\caption{
\label{fig:local_moment}
(a) Linear-log plot of the local moment at $d$-sites of the PLL as a function of temperature, for different values of $U/t$.
	(b) Same as (a), but for $p$-sites of the PLL.
	(c) Local moment on $p$ (empty symbols) and $d$ (filled symbols) sites of the PLL as a function of $U/t$, at fixed temperature $T/t = 0.1$.
	(d)–(f): Corresponding data for the LLL.
	All results correspond to lattice size $\tilde{L} = (4,4,4)$.
}
\end{figure}
%

\subsection{Magnetic orderings}

In view of the results of the previous sub-section, we start the analysis of types of magnetic orderings with a discussion on the local moment, Eq.\,\eqref{eq:local_moment}. 
Figure \ref{fig:local_moment} shows our DQMC results for the orbital-resolved local moment as functions of $T/t$ and $U/t$.
There are several important features to emphasize here. 
First, since the orbitals are not equivalent, it is observed that \(\langle m_{p}^{2} \rangle > \langle m_{d}^{2} \rangle\). 
In fact, a significant difference exists in the strength of the local moments at \(p\)- and \(d\)-sites for the PLL geometry, whereas they are quite similar for the LLL geometry. 
This difference can be traced back to the fact that the former displays a FB formed by  \(p\)-orbitals, which, in turn, are expected to host highly localized electrons. 
For the LLL geometry, on the other hand, the local moments on $p$ and $d$ orbitals display very similar temperature dependences, almost as if the data for $\ave{m_p^2}$ and $\ave{m_d^2}$ were shifted with respect to each other by a small amount; it seems that the dispersive character of the bands tends to wash out their difference. 
Second, although both \(p\) and \(d\) local moments are formed around \(T/t \sim 1-2\) and subsequently stabilize at low temperatures, there is a slight difference in the energy scales associated with their formation, especially for the PLL geometry; see Fig.\,\figref{fig:local_moment}{(c)}. This difference results in the appearance of more than one maximum in the specific heat, as shown in Fig.\,\figref{fig:SHeat}{(c)}.
%
\begin{figure}[t] 
\includegraphics[width=\linewidth]{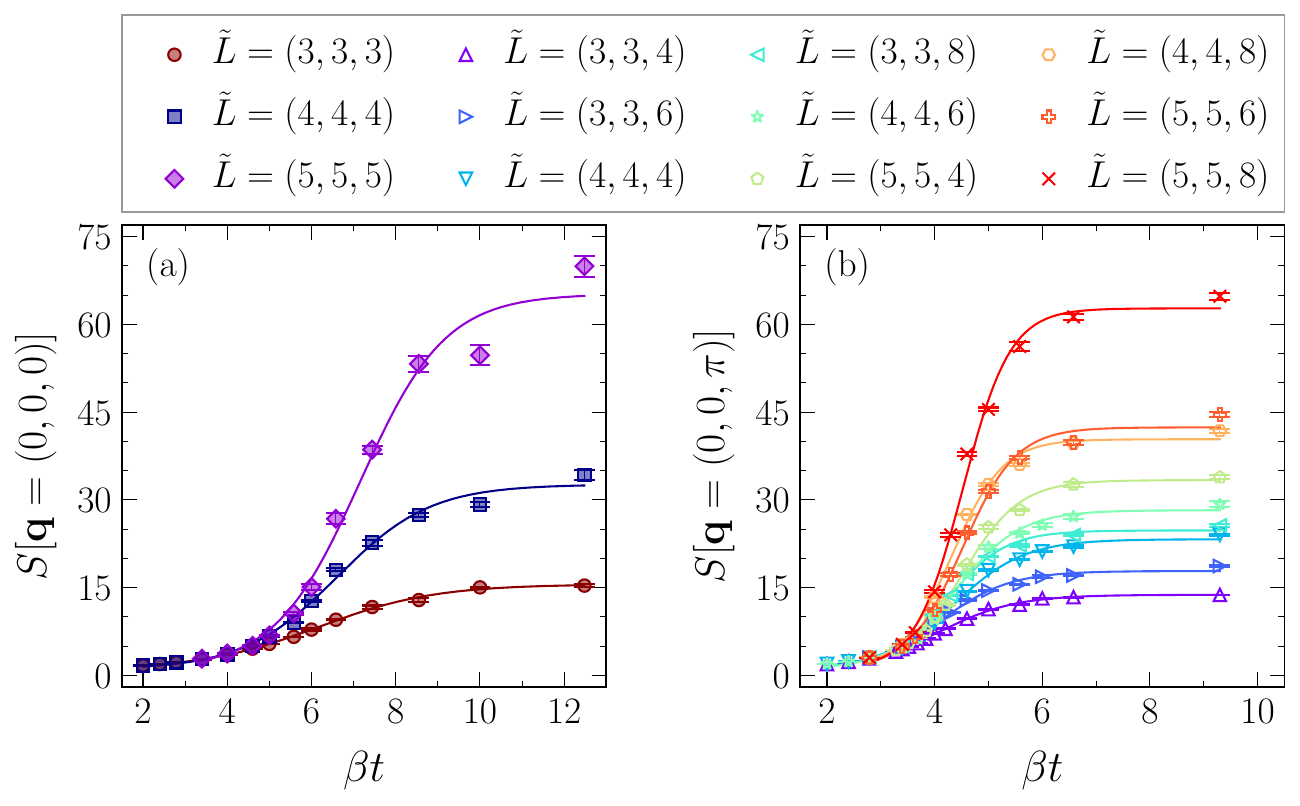}
\caption{
\label{fig:Sfactor}
Global magnetic structure factor as functions of inverse temperature for (a) the PLL and (b) the LLL. DQMC data are for different lattice sizes, $\tilde{L}$, and fixed $U/t = 5.0$; solid lines are guides to the eye.}
\end{figure}
%

Figures \figref{fig:local_moment}{(c)} and \figref{fig:local_moment}{(f)} show $\langle m_{\alpha}^{2} \rangle$ as a function of $U/t$ at $T/t=0.1$ for the PLL and the LLL, respectively. 
In both cases, the local moment increases with $U$, as a result of increasing localization of the electrons, favoring the formation of ordered states at this temperature.
Further, Fig.\,\figref{fig:local_moment}{(f)} illustrates that the small difference between $\ave{m_p^2}$ and $\ave{m_d^2}$ for the LLL extends over a wide range of values of $U$. 
The magnetic properties in strong coupling, $U/t \gg 1.0$, will be discussed below in the context of the Heisenberg model.

Let us now discuss the nature of magnetic orderings. 
Figure~\ref{fig:Sfactor} shows the dominant global magnetic structure factor for both lattices, as  functions of the inverse temperature for a given $U/t$ and different system sizes.
The saturation at low temperatures occurs as a result of the range of correlations in the ground state being limited by the finite system sizes.
For the PLL the dominant arrangement is uniform, $\bf{q}=(0,0,0)$, corresponding to ferro- or ferrimagnetism, while for the LLL the dominant arrangement, $\bf{q}=(0,0,\pi)$, corresponds to planes stacked along the $z$ direction with their magnetizations antiparallel to each other, thus leading to a global antiferromagnetic state.
We also note that the temperature range in which the structure factors display a rapid increase provides a rough guide to the critical temperature, similarly to the low-T peaks in the specific heat: according to Fig.\,\ref{fig:Sfactor}, one may expect $T_c^\text{AFM}>T_c^\text{FM}$.

\begin{figure}[t] 
\includegraphics[width=\linewidth]{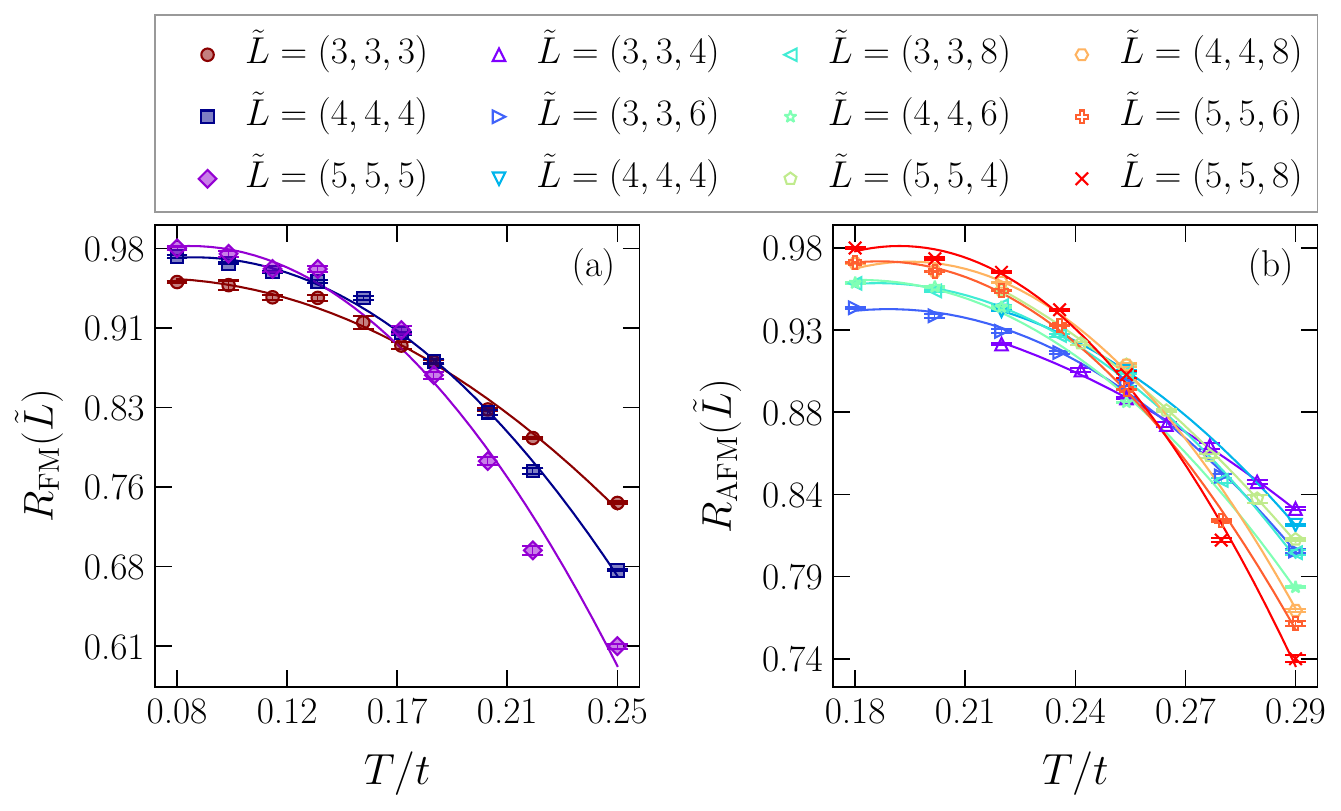}
\caption{
\label{fig:Rc}
 The magnetic correlation ratio as a function of temperature and different lattice sizes $\tilde{L}$, for (a) PLL and (b) LLL, at fixed $U/t = 5.0$. The crossing points separate the (a) paramagnetic–ferromagnetic and (b) paramagnetic–antiferromagnetic phases.
 }
\end{figure}

Accurate estimates for the critical temperatures may be obtained with the aid of the correlation ratios, Eq.\,\eqref{eq: Rc}.
In Fig.\,\ref{fig:Rc}, we illustrate \( R_\eta(\tilde{L}) \) for the corresponding dominating arrangements as a function of temperature for a fixed value of $U/t=5.0$ and various lattice sizes. 
We see that the crossings of \( R_\eta(\tilde{L}) \) occur approximately at \( T/t \approx 0.17 \), and \( T/t \approx 0.26 \), respectively for the PLL and the LLL, so that $T_c^\text{AFM}>T_c^\text{FM}$ for the same $U$, as discussed above.
We may invoke a Peierls-like argument in $\kv$-space to attribute this difference to the high degeneracy of the FB: the entropic contribution to the change in free energy is much larger for the PLL than for the LLL, so that a smaller critical temperature is needed to break the ordered state in the former case.
%
%

\subsection{Phase diagrams}
\label{subsec:phase.diagrams}
%
\begin{figure}[b] 
\includegraphics[width=\linewidth]{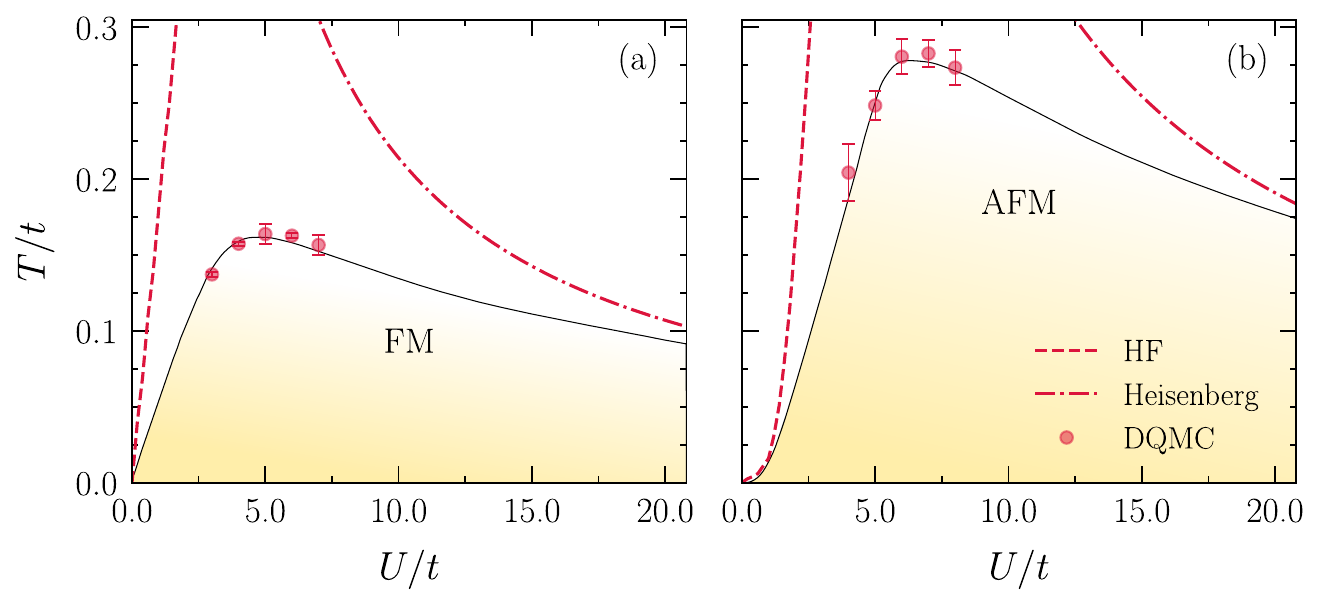}
\caption{
\label{fig:Tc}
 Magnetic phase diagram of the 3D Hubbard model on the (a) PLL and (b) LLL. Solid line are guides to the eye, created by interpolating
both the QMC results and mean-field calculations - HF and CMFT approaches.} 
\end{figure}
%

By repeating the procedure outlined above for \( R_\eta(\tilde{L}) \) for different values of \( U/t \), we set up phase diagrams  \( T_c/t \times U/t \) for both PLL and LLL geometries; in order to single out the effects of FB's we take $t_{xy}=t_z=t=1$. 
The DQMC estimates obtained from plots similar to those in Fig.\,\ref{fig:Rc} are shown as data points in Fig.\,\ref{fig:Tc} for each geometry. 
We see that AFM ordering is more resilient to the temperature than FM ordering in the corresponding lattices.
We also note that in each case the DQMC estimates for $T_c$ are able to locate the maximum $T_c$ with good accuracy. 
However, we also note that our DQMC data are restricted to ranges of $U$ near the maxima, since reaching low temperatures in three-dimensional geometries with more than one site per unit cell is computationally very costly.
In view of this, we supplement our analyses with mean-field--based approaches in weak and strong coupling regimes.

%
\begin{figure}[t] 
\includegraphics[width=\linewidth]{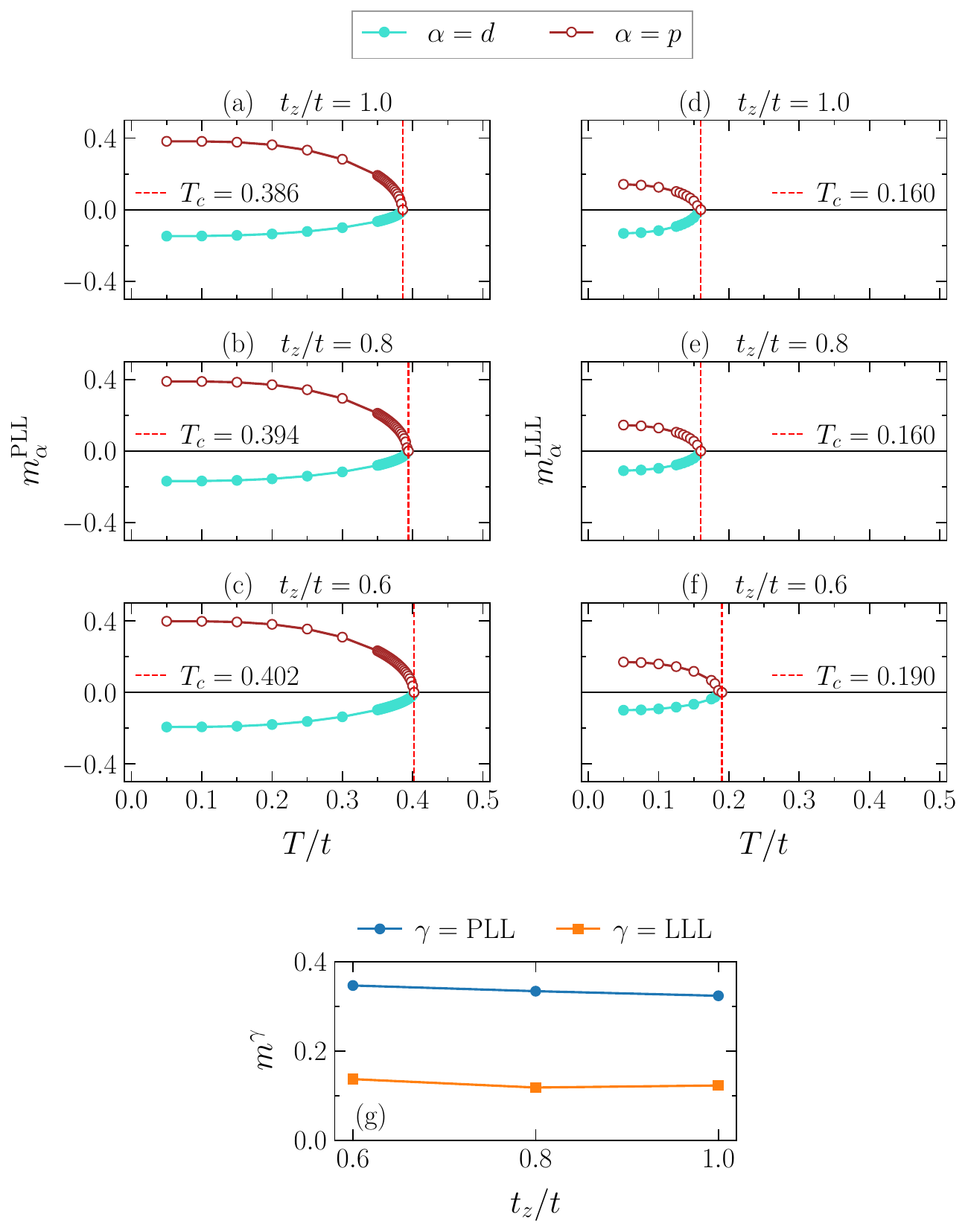}
\caption{Orbital-resolved order parameters as functions of temperature, obtained within the HF approach for different values of $t_z/t$, at fixed $U/t=2.0$. (a)-(c): PLL, and (d)-(f): LLL. 
Empty and filled symbols respectively denote data for the $p$ and $d$ orbitals, and the dashed (red) vertical lines locate $T_c$'s. (g) Order parameter as a function $t_z/t$ 
for fixed $U/t=2.0$ and $T/t=0.1$.
}
\label{fig:Tc_MFT}
\end{figure}

In weak coupling we resort to a static HF approach, the details of which are outlined in Appendix \ref{Ap}. 
We solve the equations self-consistently to obtain the magnetic order parameter as a function of temperature; see Figs.\,\figref{fig:Tc_MFT}{(a)} and \figref{fig:Tc_MFT}{(d)}, for the corresponding orderings in the PLL and LLL, respectively. 
The orbital-resolved magnetizations vanish at the same temperature, $T_c^\text{HF}$; this procedure is repeated for different values of $U$, and the resulting $U$ dependence is shown as dashed red lines in Figs.\,\figref{fig:Tc}{(a)} and \figref{fig:Tc}{(b)}.  
A clear difference in the low-temperature behavior is evident: while $T_c^\text{HF}$ vanishes linearly with $T \to 0$ in the PLL case, it vanishes exponentially in the LLL case.
This is directly related to the presence or absence of a FB: in the latter case, the dispersive bands naturally induce a competition between kinetic and potential energies, which favors itinerancy when temperature or any perturbation is introduced. 
Indeed, the behavior of $T_c^\text{HF}$ in the LLL  is similar to that of the half-filled Hubbard model on a 3D cubic lattice \cite{Hirsch87a,scalettar89}. 
On the other hand, the linear behavior of $T_c^\text{HF}$ in the PLL indicates that the system is significantly more sensitive to  \( U/t \), due to the reduced kinetic cost associated with the FB's. 
In other words, even very small values of \( U/t \) are sufficient to generate critical temperatures comparable to the peak value observed at \( U/t \approx 5.0 \).  
This behavior is in line with the trend of the magnetic order parameter  observed in the 2D LL \cite{Costa16}, and in nodal-line semimetal systems \cite{Medeiros2025}, where the magnetization rapidly saturates at its strong-coupling value.
%
\begin{figure}[t] 
\includegraphics[width=\linewidth]{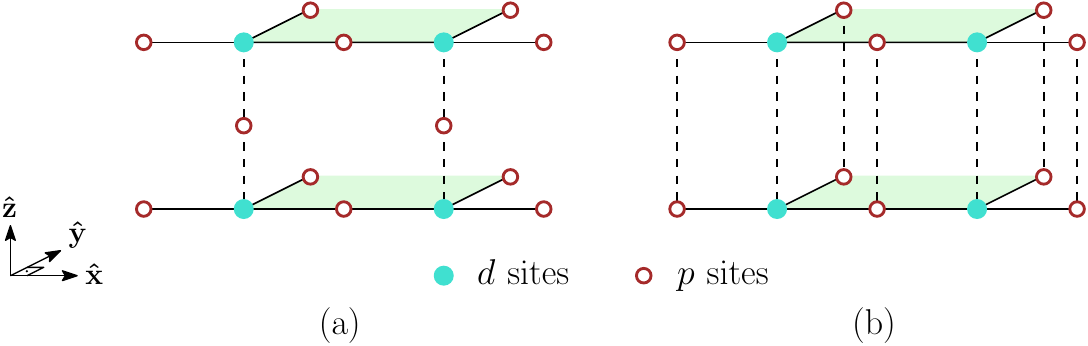}
\caption{
\label{fig:cluster_cmft}
Clusters used in our CMFT approach: (a) PLL and (b) LLL.}
\end{figure}
%

In the strong coupling regime $U/t \gg 1.0$, charge degrees of freedom are frozen and the half-filled Hubbard model is mapped onto a Heisenberg model with exchange coupling \( J \propto t^2/U \). 
Let us then examine the Heisenberg model in these geometries through a CMFT approach [see Appendix \ref{CMFT}], which amounts to solving a finite cluster of spins exactly via full diagonalization while incorporating Weiss fields at the boundaries; this procedure leads to a set of self-consistent equations for the order parameter. 
As a mean-field method, CMFT becomes more accurate for higher-dimensional systems, due to the decreasing importance of fluctuations.  
The clusters used here are illustrated in Fig.\,\ref{fig:cluster_cmft}.

\begin{figure}[t] 
\includegraphics[width=\linewidth]{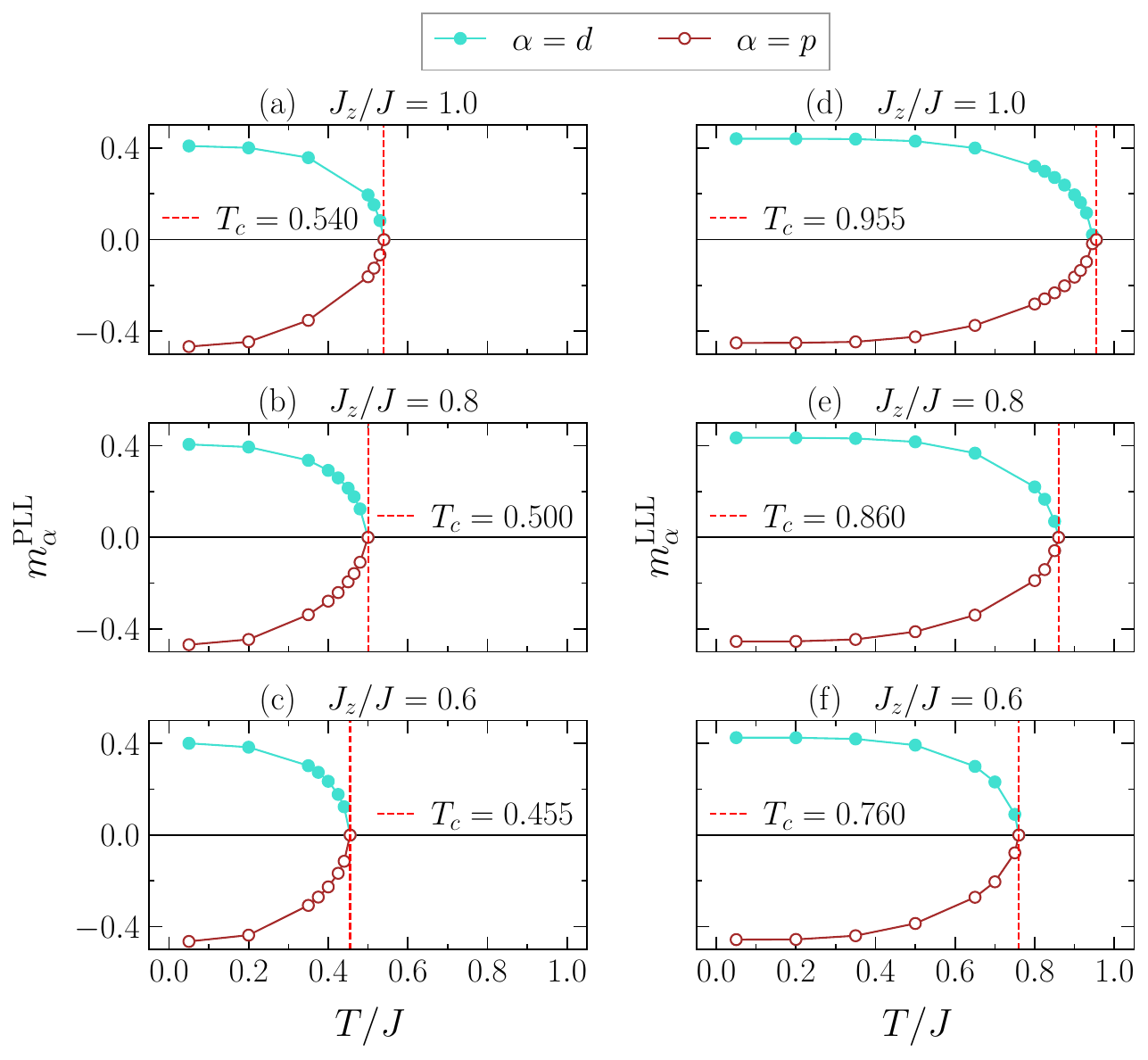}
\caption{
\label{fig:Tc_cmft}
Orbital-resolved order parameters as functions of temperature, obtained within the CMFT for different values of $J_z/J$. (a)-(c): PLL, and (d)-(f): LLL. 
Empty and filled symbols respectively denote data for the $p$ and $d$ orbitals, and the dashed (red) vertical lines locate $T_c$'s.
}
\end{figure}

The temperature dependence of the CMFT order parameters is shown in Figs.\,\figref{fig:Tc_cmft}{(a)} and  \figref{fig:Tc_cmft}{(d)} for both geometries. 
In each case, the orbital-resolved magnetizations vanish at a critical temperature, $T_c^\text{CMFT}$.
However, differently from what happens in weak coupling, now $T_c^\text{AFM} > T_c^\text{FM}$. 
We can understand this by invoking a simple mean-field argument for a uniform lattice of spins-1/2, according to which $k_\text{B}T_c = zJ/4$, so that for the case at hand we replace the coordination number by an effective one, $z\to z_\text{eff}$. 
Since $z_\text{eff}^\text{LLL}> z_\text{eff}^\text{PLL}$, the corresponding $T_c$'s follow suit.
One may also argue that hopping between four-coordinated sites assisted by a $d$-site are second-order processes, hence $J \sim (t^2/U)^2$, leading to a smaller $J_\text{eff}$; therefore, since there are more of these in the PLL than in the LLL, one expects a smaller $T_c$ in the former geometries.

We can provide somewhat better estimates for $T_c$ by recalling that, by neglecting spin fluctuations mean-field approximations overestimate $T_c$, as compared with exact or less biased numerical solutions, such as the CMFT approach.
We may then use our numerical results for $T_c^\text{CMFT}$ to express the effective coordination number as
\begin{equation}
	z_{\rm{eff}} = 4 \frac{k_B T_c^\text{CMFT}}{J}.
\end{equation}
Thus, with $z \to z_{\rm{eff}}$ and $J \to J_{\rm{eff}} = 4 t^2/U$ we may write
\begin{equation}
\frac{T_c^\text{Heis}}{t} = 4 \frac{T_c^\text{CMFT}}{J} \times \frac{t}{U}.
\end{equation} 
By using the numerically determined $T_c^\text{CMFT}$, we obtain more accurate estimates of $T_c$ in the Heisenberg limit; the results are shown as red dash-dotted lines in Fig.\,\ref{fig:Tc}.
Finally, based on estimates from the weak and strong coupling limits, along with QMC results for intermediate values of \( U/t \), we interpolate the solid black curves in Fig.\,\ref{fig:Tc} representing the critical temperatures of the Hubbard model for each geometry.

%
%
\begin{figure}[t] 
\includegraphics[width=\linewidth]{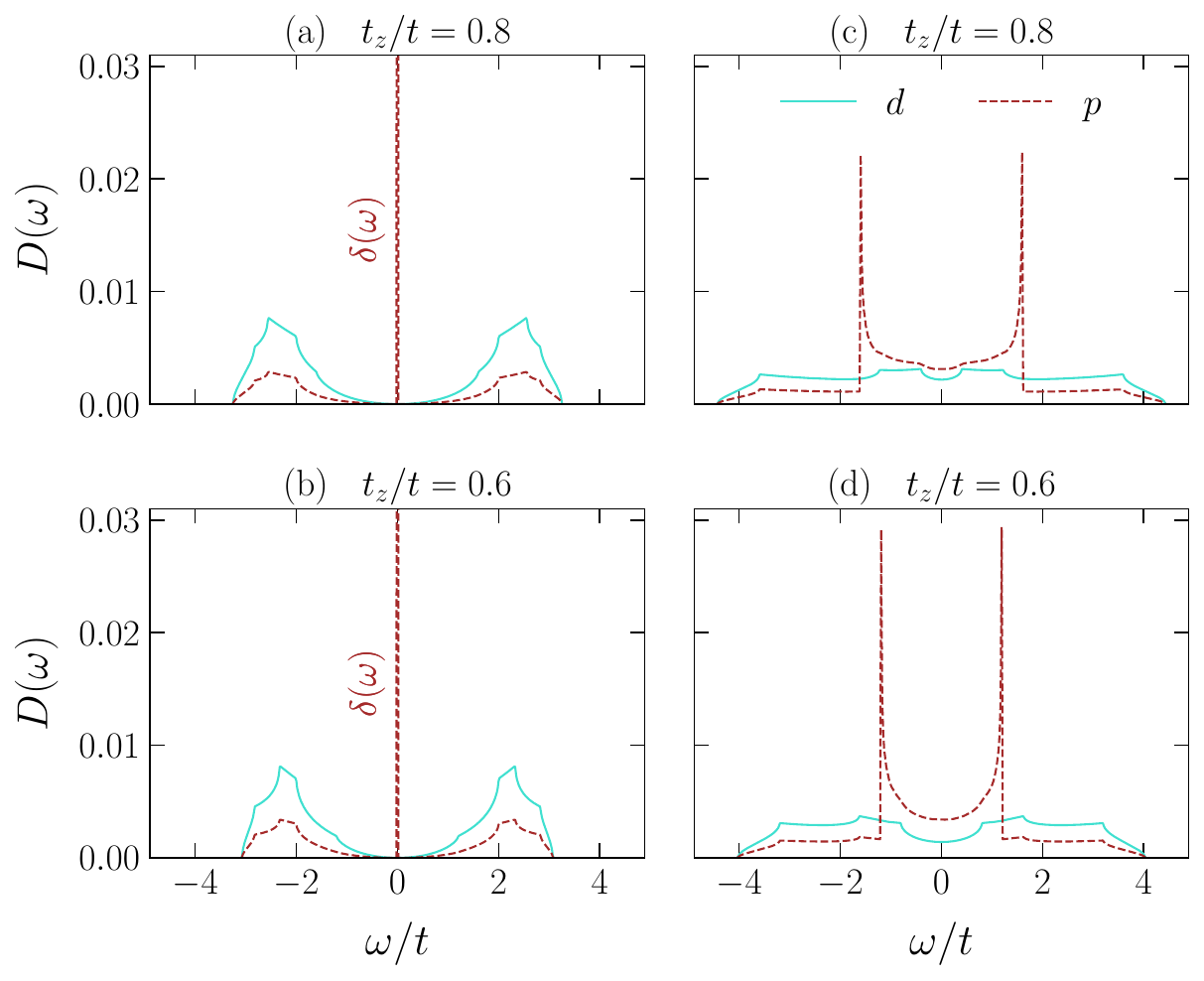}
\caption{
\label{fig:DOS_tz}
Same as Fig.\,\ref{fig:DOS}, but for different values of $t_z/t$, showing results for (a)-(b) the PLL and (c)-(d) the LLL.}
\end{figure}
%

\subsection{Effects of anisotropy}

We now briefly relax the assumption of isotropic hopping integrals by allowing for $t_z\neq t_{xy}=t$; this may describe uniaxial strain in the system. 
By the same token, we separately relax the assumption of homogeneity in $U$, by considering $U_p\neq U_d$, a situation related to atoms with distinct electronic properties occupying $p$ and $d$ sites.
In what follows, we consider each of these generalizations in turn.
With the purpose of highlighting the main qualitative differences in the magnetic responses between the LLL and the PLL, analyses will be carried out mostly within HF and CMFT approaches, in the regimes of weak and strong couplings, respectively.

\subsubsection{Anisotropy in $t_z$}

The effects of hopping anisotropy on the HF magnetization are shown in Figs.\,\figref{fig:Tc_MFT}{(a)}$-$\figref{fig:Tc_MFT}{(c)} for the PLL, and in Figs.\,\figref{fig:Tc_MFT}{(d)}$-$\figref{fig:Tc_MFT}{(f)} for the LLL, and summarized in Fig.\,\figref{fig:Tc_MFT}{(g)} by the respective absolute values of the magnetization.
We note that for both lattice geometries the critical temperatures increase with decreasing $t_z/t$, though the relative increase for the LLL is somewhat larger than for the PLL. 
Indeed, this increase in $T_c$ with decreasing $t_z/t$ for the LLL in weak coupling was also observed in the DMFT calculations for the same geometry \cite{Noda15}.
This behavior can be understood by examining the noninteracting DOS, as shown in Fig.\,\ref{fig:DOS_tz}. In the case of the PLL geometry -- panels (a) and (b) --  
anisotropy does not affect the presence of flat bands. As a result, the interacting properties of the system remain largely unchanged. 
By contrast, varying $t_z$ in the LLL geometry -- panels (c) and (d) -- alters the DOS by increasing the spectral weight associated with the $p$-orbitals while decreasing that of the $d$-orbitals. This redistribution enhances the magnetic response of the $p$-orbitals and, since these orbitals are associated with a larger number of lattice sites, the overall magnetic response of the system increases correspondingly.

At any rate, we must keep in mind that the regime $t_z\ll t$ corresponds to weakly coupled two-dimensional LL's, so that one should have $T_c\to0$ for both the PLL and the LLL, by virtue of the Mermin-Wagner theorem; that is, we expect these HF predictions to break down at some $t_z/t < 1$. 
Another interesting feature emerges by examining the orbital-resolved magnitude of the order parameter. 
While for the LLL we find $m^{\rm{LLL}}_d \sim m^{\rm{LLL}}_p$, in the PLL $m^{\rm{PLL}}_p$ is roughly twice as large as $m^{\rm{PLL}}_d$. This difference originates from the flat band character present in the PLL, which enhances the local magnetic response on $p$ sites due to the high density of states at $\epsilon_F$.
  
In the opposite limit -- namely the Heisenberg model -- the critical temperature \( T_c \) tracks the coordination number (or its effective value). Consequently, reducing the ratio \( J_z/J \) should decrease \( T_c \).
Figure \ref{fig:Tc_cmft} shows the magnetizations as a function of $J_z/J$, from which we see that the critical temperature decreases with decreasing $J_z/J$ for both geometries, although faster for the LLL than for the PLL.
This larger sensitivity in the LLL can be attributed to its geometry, which includes a larger number of links along the \( z \)-direction. As a result, the effective coordination number in the LLL is more strongly influenced by changes in \( J_z/J \).

Based on these two limiting cases, we can infer some expectations for the finite-temperature phase diagrams in the presence of anisotropy along the \(z\)-direction.  
In the PLL case, since both the weak- and strong-coupling limits are weakly affected by anisotropic effects, small variations in the ratio \(t_z/t\) are expected to produce a phase diagram very similar to that shown in Fig.\,\figref{fig:Tc}{(a)}.  
By contrast, the LLL geometry exhibits an enhancement of \(T_c/t\) in the weak-coupling regime and a suppression in the strong-coupling regime. This behavior could effectively shift the maximum of \(T_c/t\) in Fig.\,\figref{fig:Tc}{(b)} to lower values of \(U/t\).

\subsubsection{Inhomogeneity in $U$}

We now explore some consequences of allowing for inhomogeneous on-site couplings, i.e.,  \( U_d/t \neq U_p/t\).  
First, recall that in strong coupling, \( U_d \neq U_p \gg t \), it is still possible to map the LL onto an effective Heisenberg model with exchange coupling $J' = 4t^2/ \widetilde{U}$, where \( \widetilde{U} \) is the geometric mean of the on-site repulsions between neighboring sites~\cite{Franca10}; hence a 2D LL sustains FM order in the ground state, as long as $U_p/t,\, U_d/t\neq0$ \cite{Costa16}.
However, when either \( U_p/t = 0 \) or \( U_d/t = 0 \), this mapping breaks down and Lieb's theorem cannot be invoked. 
Nonetheless, DQMC simulations provided numerical evidence supporting a FM ground state when $U_p/t\neq0$ and $U_d/t=0$, but not the other way around \cite{Costa16}. 
For completeness, we should mention that ferrimagnetism is sustained even when approximately half of the lattice sites are randomly assigned a nonzero interaction \( U \)\,\cite{lima20}, a fraction which exceeds both the classical and quantum percolation thresholds for the 2D LL\,\cite{Oliveira21,Oliveira25}.

Let us then discuss how these findings affect the 3D extensions of the LL. 
First, as long as $U_p/t,\, U_d/t\neq0$, the FM planes remain coupled along the $z$ direction, giving rise to the FM (PLL) and AFM (LLL) arrangements, similarly to the homogeneous case. 
When $U_p/t\neq0$ and $U_d/t=0$ (not shown), the three orthogonal FM planes on the PLL can accommodate a global FM state; for the LLL, on the other hand, the FM planes are antiferromagnetically coupled through the $p$ sites, thus giving rise to a layered AFM state.

%
\begin{figure}[t] 
\includegraphics[width=\linewidth]{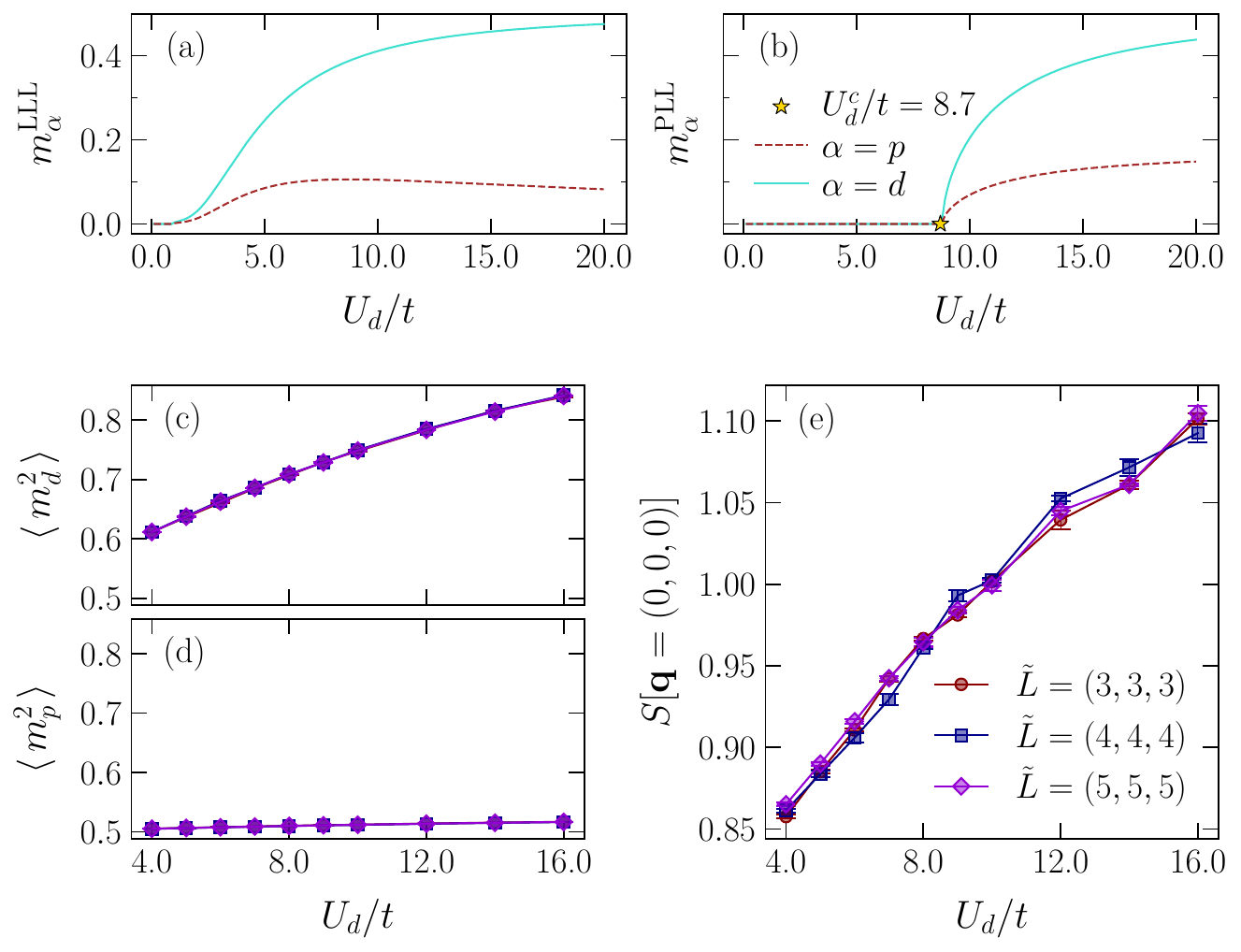}
\caption{
(a) Orbital-resolved HF ground-state magnetizations as functions of $ U_d/t $, for the LLL with $ U_p/t = 0 $.  
(b) Same as (a), but for the PLL, where the critical value $U^c_d/t$ is indicated; see text.
(c) DQMC results for local moments at $d$ orbitals as functions of $ U_d/t $, for the LLL with $ U_p/t = 0 $, $\beta =16/t$, and lattice sizes $\tilde{L}$ indicated in (e).
(d) Same as (c), but at $p$ orbitals.
(e) Same as (c) but for the total magnetic structure factor.}
\label{fig:fig_mgz_m2_and_SF_vs_U}
\end{figure}
%

We are therefore left with the more subtle case of $U_p/t=0$ and $U_d/t\neq0$.
Starting with the LLL, the system effectively consists of one-dimensional chains, each of which with interactions driven by $U_d/t$, and weakly coupled through horizontal hoppings via the $p$-orbitals. 
Through the HF approximation, we obtain the ground state magnetizations as functions of $U_d/t$, as shown in Fig.\,\figref{fig:fig_mgz_m2_and_SF_vs_U}{(a)}.
As expected, we observe that the magnetization at the $d$-orbitals is significantly larger than at the $p$-sites, which agrees with our claim that, in this regime, the system consists of weakly coupled one-dimensional chains of $d$-orbitals. Additionally, both order parameters exhibit an exponential dependence as $U_d \to 0$, a hallmark of mean-field solutions in systems with a finite density of states at the Fermi level.
That is, the $U_p/t=0$  LLL is magnetic for any \( U_d/t > 0 \). We emphasize that this picture of weakly coupled 1D chains is supported by DQMC simulations on Hubbard model superlattices \cite{Mondaini17}.

In stark contrast, the mean-field magnetic response for the PLL with $U_p/t=0$ and $U_d/t\neq0$ is shown in Fig.\,\figref{fig:fig_mgz_m2_and_SF_vs_U}{(b)}: it 
displays a quantum critical point (QCP) at \( U^c_d/t\approx 8.7 \), separating a nonmagnetic ground state from a FM one.
The possibility of realizing a QCP, even within a mean-field framework, motivated us to explore this further using DQMC simulations.
In order to check this, Figs.\,\figref{fig:fig_mgz_m2_and_SF_vs_U}{(c)} and \figref{fig:fig_mgz_m2_and_SF_vs_U}{(d)} show the DQMC results for the local moment on $d$ and $p$ sites, respectively.
While $\ave{m_d^2}$ increases steadily with $U_d/t$, indicating increasingly localized moments, $\ave{m_p^2}$ remains small and close to the non-interacting limit, indicating that these sites do not support significant magnetic ordering.
The spin structure factor offers a more stringent test of long-range magnetic ordering: indeed, the DQMC data for $S[\bf{q} = (0,0,0)]$ in Fig.\,\figref{fig:fig_mgz_m2_and_SF_vs_U}{(d)}, show no increase with $\tilde{L}$, indicating that magnetic correlations are short-ranged, instead of long-ranged. 
The inescapable conclusion is that the presence of a QCP in the $U_p/t=0$ PLL appears to be an artifact of the mean-field approximation.
  
\section{Conclusions}
\label{sec:conclusions}

We have investigated how the magnetic properties of a strongly correlated two-dimensional system with a flat band, namely the half-filled Hubbard model on a Lieb lattice (LL), are changed by stacking into three-dimensional structures. 
The interest stems from the fact that stacking planes exhibiting a ferrimagnetic state in the ground state could lead to global ferrimagnetism or antiferromagnetism, depending on how the stacking is carried out.
In addition, three-dimensional structures can display magnetic order at finite temperatures, so that the interplay between the high degeneracy of flat bands and the temperature could lead to interesting features.
We have focused on two geometries, the perovskite Lieb lattice (PLL) and the layered Lieb lattice (LLL) [see Fig.\,\ref{fig:lattice}], since the former preserves a flat band and the latter suppresses it.
 
The picture that emerges from our DQMC simulations at half filling is that the PLL orders ferromagnetically as a result of the preserved cubic symmetry, while the LLL orders antiferromagnetically, in a sequence of oppositely magnetized planes. 
In each case, the critical temperature displays a maximum as a function of the on-site repulsion, $U$, with  $T_c$'s for the PLL generally lying below those for the LLL; this is attributed to the high degeneracy of the flat band present in the former, which is absent in the latter.

We have also resorted to Hartree-Fock (HF) and cluster--mean-field theory (CMFT) approaches, respectively in weak- and strong couplings, to supplement our DQMC simulations. 
For instance, these calculations suggest that $T_c$ vanishes linearly and exponentially as $U\to0$, for the PLL and the LLL, respectively.
In strong-coupling, both systems are described by an effective Heisenberg model, with $T_c \propto z_\text{eff}J$; thus, since the effective number of neighbors is smaller for the PLL than for the LLL, the asymptotic (i.e., $U/t \gg 1.0$) $T_c$ curve for the former lies below the one for the latter. 

We have also used these mean-field approaches to investigate the effect of hopping and exchange anisotropies along the stacking direction. 
Our results suggest that the maxima in $T_c(U)$ should shift towards lower values of $U/t$, for decreasing anisotropies.
We have also considered the possibility of having different values of on-site repulsion, $U_p\neq U_d$. 
When $U_p/t=0$ and $U_d/t\neq0$, our HF calculations suggest the existence of a quantum critical point for the PLL, which was not confirmed by DQMC simulations.
The other possibility, $U_p/t\neq0$ and $U_d/t=0$ did not yield surprising results. 

In closing, the fact that the high degeneracy of flat bands can affect the critical temperature for magnetic order may indicate that other finite temperature phase transitions could undergo significant changes in scenarios with flat bands.

\section*{Acknowledgments}

The authors are grateful to the Brazilian Agencies Conselho Nacional de Desenvolvimento Científico e Tecnológico (CNPq), Coordenação de Aperfeiçoamento de Pessoal de Ensino Superior (CAPES)
for funding this project. 
J.F. acknowledges support from FAPERJ grant No.\,260003/019642/2022 and ANID
Fondecyt grant No.\,3240320. 
N.C.C. also acknowledges support from FAPERJ grant No.\,E-26/200.258/2023 - SEI-260003/000623/2023, and CNPq grant No.\,313065/2021-7. 
R.R.d.S. acknowledges support from FAPERJ grant No.\,E-26/210.974/2024, and CNPq grant No.\,314611/2023-1. This research was
partially supported by the supercomputing infrastructure
of the NLHPC (CCSS210001).

\appendix

\section{Mean-field Hartree-Fock approximation}
\label{Ap}

Within the HF approximation, we decouple the quartic interaction terms into a quadratic form through mean-field decomposition~\cite{bruus04}. Disregarding the terms that do not conserve the number of particles, $\hat{n}^{\alpha}_{\rv, \uparrow} \hat{n}^{\alpha}_{\rv, \downarrow}$ is decoupled as follows:
\begin{align}
\label{eq:Hubbard_decoupled}
\alpha_{\rv \uparrow}^{\dagger} \alpha_{\rv \uparrow}\langle \alpha_{\rv \downarrow}^{\dagger} \alpha_{\rv \downarrow}\rangle + \langle \alpha_{\rv \uparrow}^{\dagger} \alpha_{\rv \uparrow}\rangle \alpha_{\rv \downarrow}^{\dagger} \alpha_{\rv \downarrow} - \langle \alpha_{\rv \uparrow}^{\dagger} \alpha_{\rv \uparrow} \rangle \langle \alpha_{\rv \downarrow}^{\dagger} \alpha_{\rv \downarrow} \rangle \,+ \nonumber \\
-\alpha_{\rv \downarrow}^{\dagger} \alpha_{\rv \uparrow} \langle \alpha_{\rv \uparrow}^{\dagger} \alpha_{\rv \downarrow} \rangle - \langle \alpha_{\rv \downarrow}^{\dagger} \alpha_{\rv \uparrow} \rangle \alpha_{\rv \uparrow}^{\dagger}  \alpha_{\rv \downarrow} + \langle \alpha_{\rv \uparrow}^{\dagger} \alpha_{\rv \downarrow} \rangle \langle \alpha_{\rv \downarrow}^{\dagger} \alpha_{\rv \uparrow} \rangle\, ,
\end{align}
where $\alpha = d$, $p^x$, $p^y$ or $p^z$ orbitals. Combining spin operators in a fermionic basis,
\begin{equation}\label{eq:spin-operators}
	\rb{\hat{S}_r}^\alpha = \tfrac{1}{2} \sum_{s,\,s^\prime=\pm} \alpha^{\dagger}_{\,\rb{r}, s} \bm{\hat{\sigma}}_{s s^\prime} \alpha_{\,\rb{r}, s^\prime},
\end{equation}
with $\bm{\hat{\sigma}}_{s s^\prime}$ denoting Pauli matrix elements, and $\hat{n}^{\alpha}_{\rv} = \hat{n}^{\alpha}_{\rv \uparrow} + \hat{n}^{\alpha}_{\rv \downarrow}$ for the respective orbitals, Eq.\,\eqref{eq:Hubbard_decoupled} is written as
\begin{align}
	\frac{\ave{\hat{n}^{\alpha}_{\rv}}}{2}\hat{n}^{\alpha}_{\rv} -\frac{\ave{\hat{n}^{\alpha}_{\rv}}^2}{4} - 2 \, \ave{\bf{\hat{S}^{\alpha}}_\rb{r}} \cdot \bf{\hat{S}^{\alpha}}_\rb{r} + \ave{\bf{\hat{S}^{\alpha}}_{\rv}}^{2}.  
\end{align}

Defining the average electron density, $\ave{\hat{n}^{\alpha}_\rb{r}} \equiv n^{\alpha}_\rb{r}$, and the average magnetization, $\ave{\rb{\hat{S}^{\alpha}_r}} \equiv \bf{m^{\alpha}_r} = (0,0,m_\alpha)$, we obtain a mean-field HF approximation in the Hubbard Hamiltonian on the PLL and LLL geometries, read as
\begin{equation}\label{eq:Hamiltonian_MF}
	\widehat{\mathcal{H}}_\rm{HF} = \widehat{H}_K + \widehat{H}_{U_\rm{HF}},
\end{equation}
where
\begin{equation}\label{eq:U-term_HF}
\widehat{H}_{U_\rm{HF}} = \left[\frac{n^{\alpha}_{\rb{r}}}{2}\hat{n}^{\alpha}_{\rb{r}} -\frac{(n^{\alpha}_{\rb{r}})^2}{4} - 2 \, \bf{m}^{\alpha}_{\rb{r}} \cdot \rb{\hat{S}^{\alpha}_r} + (\bf{m}^{\alpha}_{\rb{r}})^{2} \right]. 
\end{equation}
Performing a discrete Fourier transform and using a Nambu spinor basis, $\Psi^\dagger_\kv = [ \alpha^\dagger_{\kv, \sigma} ]$, we can write
\begin{equation}
\widehat{H}_{U_\rm{HF}} = \sum_{\kv,\alpha}\Psi^\dagger_{\kv,\sigma}\,
\mathsf{H}^\gamma \,
\Psi_{\kv,\alpha} - \sum_{\alpha} U_\alpha\left[\frac{(n^\alpha)^{2}}{4} - m^{2}_\alpha\right],
\end{equation}
where
\begin{equation}
	\label{eq:matrix_HF}
	\mathsf{H}^\gamma =  \begin{pmatrix} H_\uparrow^\gamma(\kv) & 0 \\ 0 & H_\downarrow^\gamma(\kv) \end{pmatrix}. 
\end{equation}
\begin{widetext}
\noindent
Thus, for each specific geometry, we have
 \begin{equation}
	H^\rm{PLL}_{\uparrow}(\kv) =  \begin{pmatrix} U_{d}(\frac{1}{2}n^{d}_{\uparrow}-m_{d}) & -t_{xy}(1+e^{-ik_x}) & -t_{xy}(1+e^{-ik_y}) &  -t_{z}(1+e^{-ik_z})  \\ -t_{xy}(1+e^{ik_x})  & U_{p}(\frac{1}{2}n^{p}_{\uparrow}-m_{p}) & 0 & 0\\ -t_{xy}(1+e^{ik_y}) & 0 & U_{p}(\frac{1}{2}n^{p}_{\uparrow}-m_{p}) & 0 \\ 
		-t_{z}(1+e^{ik_z}) & 0 & 0 & U_{p}(\frac{1}{2}n^{p}_{\uparrow}-m_{p})\\
	\end{pmatrix} 
\end{equation}
and
 \begin{equation}
	H^\rm{LLL}_{\uparrow}(\kv) =  \begin{pmatrix} U_{d}(\frac{1}{2}n^{d}_{\uparrow}-m_{d}) -2t_{z}\cos k_{z}& -t_{xy}( 1+ e^{-ik_x}) & -t_{xy}( 1+ e^{-ik_y})  \\ -t_{xy}( 1+ e^{ik_x})  & U_{p}(\frac{1}{2}n^{p}_{\uparrow}-m_{p}) -2t_{z}\cos k_{z} & 0 \\ -t_{xy}( 1+ e^{ik_y})  & 0 & U_{p}(\frac{1}{2}n^{p}_{\uparrow}-m_{p}) -2t_{z}\cos k_{z} \\ 
	\end{pmatrix}.
\end{equation}

\end{widetext}

Diagonalizing the matrix $\mathsf{H}^\gamma$ in Eq.~\eqref{eq:matrix_HF} yields the eigenvalues $\lambda^{(\nu)}_{\mathbf{k}}$ (with $\nu$ labeling quasiparticle bands in each geometry) and the Helmholtz free energy takes the form
\begin{equation}\label{eq:Helmholtz-free-energy}
	\mathcal{F} = - k_B T \sum_{\kv,\nu} \ln \left[ 1 + \exp\left( - \frac{\lambda^{(\nu)}_\rb{k}}{k_B T} \right)\right] - \sum_{\alpha} U_\alpha\left[\frac{(n^\alpha)^{2}}{4} - m^{2}_\alpha\right].
\end{equation}
Then, the fields $m_\alpha$ are determined self-consistently through the minimization of the Helmholtz free energy with the aid of the Hellmann-Feynman theorem,
\begin{equation}\label{eq:Hellmann-Feynman-theorem}
	\left\langle \frac{\partial\mathcal{F}}{\partial m_\alpha} \right\rangle = 0.
\end{equation}
%


\section{Cluster Mean-Field approximation: spin-$1/2$ Heisenberg model}
\label{CMFT}

The spin-$1/2$ Heisenberg model is described by the Hamiltonian
\begin{equation}
	\widehat{H} = \sum_{\langle i,j \rangle} \left[ J_{xy} \left(S^+_i S^-_j + S^+_j S^-_i\right) + J_z S^z_i S^z_j \right],
\end{equation}
where the sum runs over nearest neighbors on a $d$-dimensional lattice, and $ S^{\pm}_i $ denote the spin raising and lowering operators at site $ i $. The parameters $ J_{xy} $ and $ J_z $ correspond to the coupling constants in the transverse and longitudinal directions, respectively.

In this work, we restrict ourselves to the isotropic limit, i.e., $ J_{xy} = J_z = J $, unless stated otherwise. Within the CMFT framework, we treat interactions inside a cluster exactly, while approximating the coupling between cluster and environment via a mean-field decoupling. Specifically, for a boundary interaction between a site $ i $ inside the cluster and a neighboring site $ i' $ outside, we perform the following replacement:
\begin{equation}
	J \, \mathbf{S}_i \cdot \mathbf{S}_{i'} \approx J \left( S^z_i \langle S^z_{i'} \rangle + \langle S^z_i \rangle S^z_{i'} - \langle S^z_i \rangle \langle S^z_{i'} \rangle \right),
\end{equation}
where we have broken SU(2) symmetry by selecting the longitudinal $ z $-component as the direction of magnetic ordering, which defines our order parameter.

On bipartite lattices, such as the square lattice, this leads to a cluster Hamiltonian of the form
\begin{multline}
	\widehat{H}_{\text{CMFT}} = \sum_{\langle i,j \rangle \in C} J \left(S^+_i S^-_j + S^+_j S^-_i + S^z_i S^z_j \right) \\
	+ J \sum_{i \,\in \,C \,\cap \,A} m_B S^z_i + J \sum_{i \,\in \,C \,\cap \, B} m_A S^z_i,
\end{multline}
where the first sum runs over all bonds within the cluster $ C $, assuming open boundary conditions. The second and third terms represent the mean-field interactions across the boundary, where sites on the edge of the cluster $ C $ belong to sublattices $ A $ or $ B $. The mean fields are defined as $ m_\alpha = \langle S^z_{i_\alpha} \rangle $, with $ \alpha = A, B $.
In the case of a bipartite square lattice with no external field, we expect antiferromagnetic ordering such that $ m_A = -m_B $. However, this relation may break down for more general lattice geometries or in the presence of a finite external magnetic field.

The local magnetizations $ m_\alpha $ are computed self-consistently from the thermal averages within each cluster configuration:
\begin{equation}
	m_\alpha = \frac{1}{\mathcal{N}_C} \sum_{n=1}^{\mathcal{N}_C} \frac{1}{n_\alpha} \sum_{j_\alpha \in C_n} \frac{\Tr\left(S^z_{j_\alpha} e^{-\beta \widehat{H}_{C_n}} \right)}{\Tr\left(e^{-\beta \widehat{H}_{C_n}} \right)},
\end{equation}
where $ \beta = 1/T $, $ n_\alpha $ is the number of sites in sublattice $ \alpha $ within cluster $ C_n $, and $ \mathcal{N}_C $ denotes the number of distinct cluster used in the average. Depending on the size of the cluster, we can choose more than one pattern for the sublattices. Because of this, we may also average over the contribution of all possible choices of clusters $\mathcal{N}_C$.

\bibliography{refs_lieb3d.bib}

\end{document}